\documentclass[prl,aps,superscriptaddress,showpacs,twocolumn,notitlepage,groupedaddress]{revtex4-1}
\usepackage{bm}
\usepackage{amsfonts}
\usepackage[dvips]{graphicx}
\usepackage{mathrsfs}
\usepackage[intlimits]{amsmath}
\usepackage[colorlinks, citecolor=red]{hyperref}
\usepackage[english]{babel}

\hypersetup{colorlinks,
linkcolor=blue,          citecolor=blue,        filecolor=blue,      urlcolor=blue           }
\graphicspath{{../figures/}}

\newcommand{\ket}[1]{| #1 \rangle}

\newcommand{\overbar}[1]{\mkern 3mu\overline{\mkern-3mu#1\mkern-3mu}\mkern 3mu}

\begin{document}

\title{Certification of Boson Sampling Devices with Coarse-Grained Measurements}

\author{Sheng-Tao Wang}
\author{L.-M. Duan}
\affiliation{Department of Physics, University of Michigan, Ann Arbor, Michigan 48109, USA}
\date{\today }

\begin{abstract}
A boson sampling device could efficiently sample from the output probability distribution of noninteracting bosons undergoing many-body interference. This problem is not only classically intractable, but its solution is also believed to be classically unverifiable. Hence, a major difficulty in experiment is to ensure a boson sampling device performs correctly. We present an experimental friendly scheme to extract useful and robust information from the quantum boson samplers based on coarse-grained measurements. The procedure can be applied to certify the equivalence of boson sampling devices while ruling out alternative fraudulent devices. We perform numerical simulations to demonstrate the feasibility of the method and consider the effects of realistic noise. Our approach is expected to be generally applicable to other many-body certification tasks beyond the boson sampling problem. 
\end{abstract}

\maketitle

In the last three decades, quantum computation has stimulated considerable excitement among physicists, computer scientists, and mathematicians, with the general belief that quantum computers could solve certain tasks significantly faster than current electronic computers \cite{nielsen2010quantum, Aaronson2013quantum}. Especially after the discovery of Shor's factoring algorithm \cite{Shor1997Polynomial}, theorists and experimentalists have been teaming up to converge on the goal of demonstrating quantum supremacy \cite{Ladd2010Quantum}. Recently, an important step made by Aaronson and Arkhipov \cite{Aaronson2011Computational} is to formulate the boson sampling problem, which is intractable for classical computers but remarkably amenable to quantum experiments. A number of elegant experiments have since implemented the problem with linear optics on a small scale \cite{Broome2013Photonic, Spring2013Boson,Tillmann2013Experimental, Crespi2013Integrated,Spagnolo2013General, Spagnolo2014Experimental, Carolan2014On, Bentivegna2015Experimental}. Pushing the experiments beyond classical capabilities would constitute a strong demonstration of the quantum speedup and in addition lead to important implications in the foundations of computer science \cite{Aaronson2013quantum, Aaronson2011Computational}. 

At the core of the hardness-of-simulation property lie the exponential cost of computing a matrix permanent \cite{Valiant1979Complexity} and the exponential number of possible output events in boson sampling. These pose major difficulties in certifying the correctness of a boson sampler on a large scale \cite{Gogolin2013Boson, Aaronson2014Bosonsampling, Spagnolo2014Experimental, Carolan2014On, Tichy2014Stringent, Aolita2015Reliable}. The credibility of a certification process thus relies on gathering convincing circumstantial evidence while ruling out alternative explanations. Several efficient schemes have been proposed to validate a boson sampler against the uniform sampler \cite{Gogolin2013Boson} making use of the information in the unitary process \cite{Aaronson2014Bosonsampling, Spagnolo2014Experimental, Carolan2014On} and against the classical sampler of distinguishable particles \cite{Broome2013Photonic, Spagnolo2014Experimental, Tillmann2015Generalized} exploiting the bosonic clouding behavior \cite{Carolan2014On}. It is also possible to depart from the computationally hard space and design an efficient test based on predictable forbidden events with special inputs and scattering process \cite{Tichy2014Stringent, Tichy2014Interference, Shchesnovich2015Partial}, assuming the device would be equally operational in general. An unsettled problem especially pertinent to experiments is whether one would be able to extract useful and robust information from a large-scale boson sampler. In other words, would any filtered information be able to verify the equivalence of two identical boson sampling devices while excluding possible fraudulent ones? In light of rapid experimental advances, this issue will be increasingly relevant when the system scales up: as the probabilities of generic output events become exponentially small, sampling noise due to limited measurement trials may conceal any distinctive information. 

In this paper, we introduce an experimental friendly scheme to extract useful structures from a boson sampling device based on coarse-grained measurements. Using standard statistical tools, we simulate the experimental certification process and show that the coarse-grained information is able to provide a quantitative assessment to the degrees of matching between two alleged boson samplers. This is important when one needs to verify the equivalence of two quantum samples drawn from the same boson sampling device or from different devices with identical processes. It will also be crucial in situations wherein we can completely trust one device and need to validate another possibly fraudulent device against the reliable one. Our numerical simulation in addition demonstrates our scheme could tolerate a moderate amount of experimental noise \cite{Rohde2012Error, Shchesnovich2014Sufficient} while strong noise invalidates the equivalence due to mismatched interference processes. On a broader scale, our method is not specific to boson sampling, but could be applicable to other generic many-body certification problems. 

\begin{figure}[t]
\hspace{-.2cm}
\includegraphics[trim=0cm 0cm 0cm 0cm, clip,width=\columnwidth]{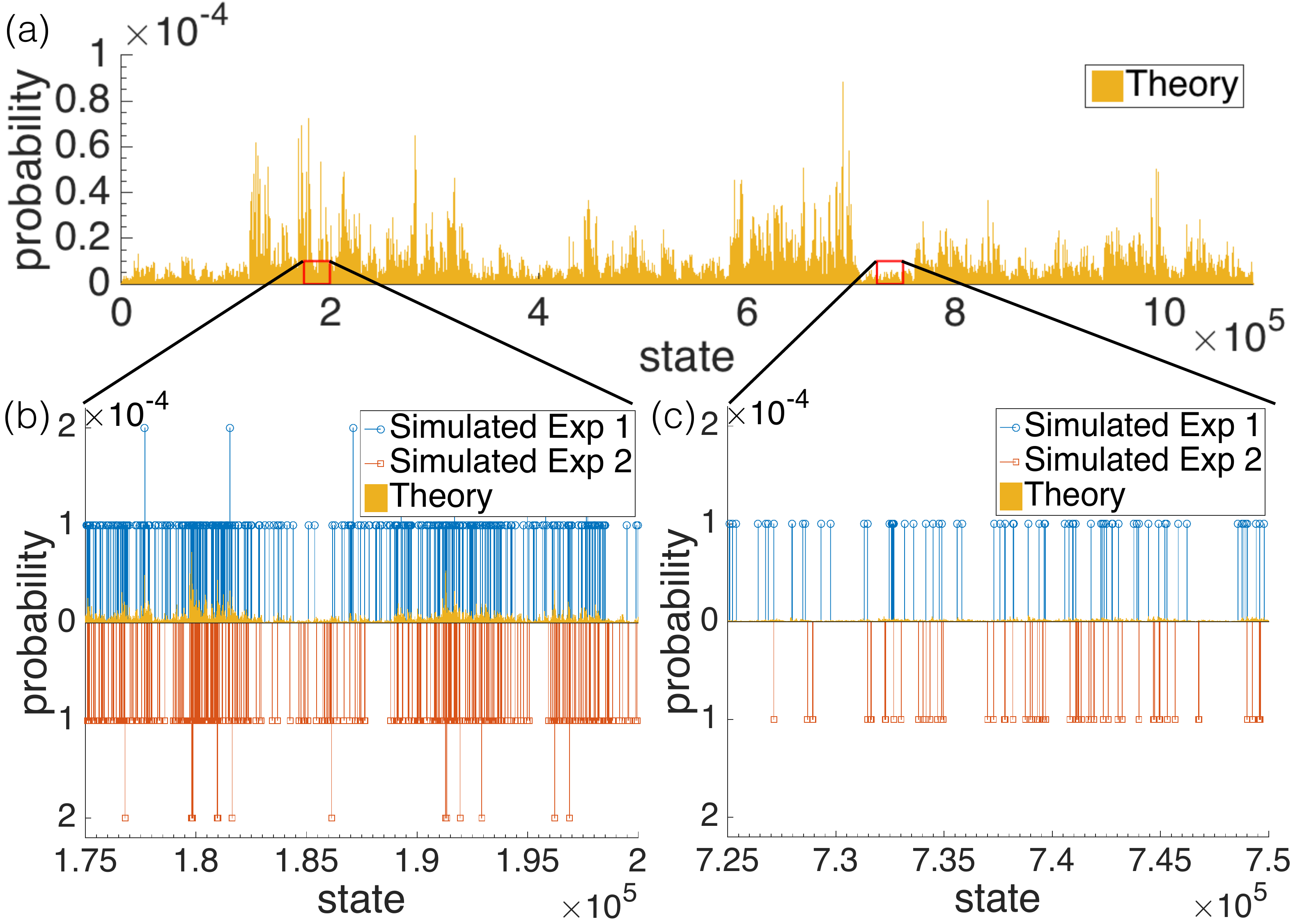} 
\caption{Original Distributions. (a) Theoretical distribution $P^{Q}$ of a boson sampling device with $N=5$ particles in $M=40$ modes after a random unitary transformation. (b) and (c) shows the zoom-in distributions of two simulated experimental samples ($P^{Q}_{S1}$ and $P^{Q}_{S2}$) with sample size $N_{m} = 10000$ drawn from $P^{Q}$. The theoretical distribution is superimposed onto the simulated samples. 
}
\label{Fig:randUOrigin}
\end{figure}

Before proceeding to our proposed scheme, we briefly introduce the boson sampling problem and clarify why large sampling errors are involved without coarse graining. In a typical setup, there are $N$ indistinguishable bosons prepared in $M$ input modes and allowed to coherently interfere with one another in a (random) unitary process. We abstract away from interactions between particles, so the resultant many-body interference is purely due to the bosonic statistics. To compute the probability of an output event, one needs to calculate the permanent of the associated $N \times N$ matrix \cite{Broome2013Photonic, Scheel2004Permanents}, which requires an exponential cost for a generic complex matrix. The possible number of output events is $D=\binom{M+N-1}{N}$, which grows exponentially with $N$ (for $M \geq N$). As the system scales up, the number of measurement runs will be unable to keep pace with the exponential growth. This gives rise to large sampling errors with limited sample size. Fig.~\ref{Fig:randUOrigin} shows an example in which the sample size is less than $1\%$ of the Hilbert space dimension. Two samples drawn from the same device could be rather dissimilar (with a low average fidelity $F \approx 0.039 \pm 0.002$, where $F=\big| \sqrt{ \smash[b]{P^{Q}_{S1}} } \cdot \sqrt{\smash[b]{P^{Q}_{S2}} } \big|$). From these distributions, it is not quite possible to assess whether the samples are drawn from the same bona fide boson sampling device.  

\begin{figure}[t]
\includegraphics[trim=0cm 0cm 0cm 0cm, clip,width=0.9\columnwidth]{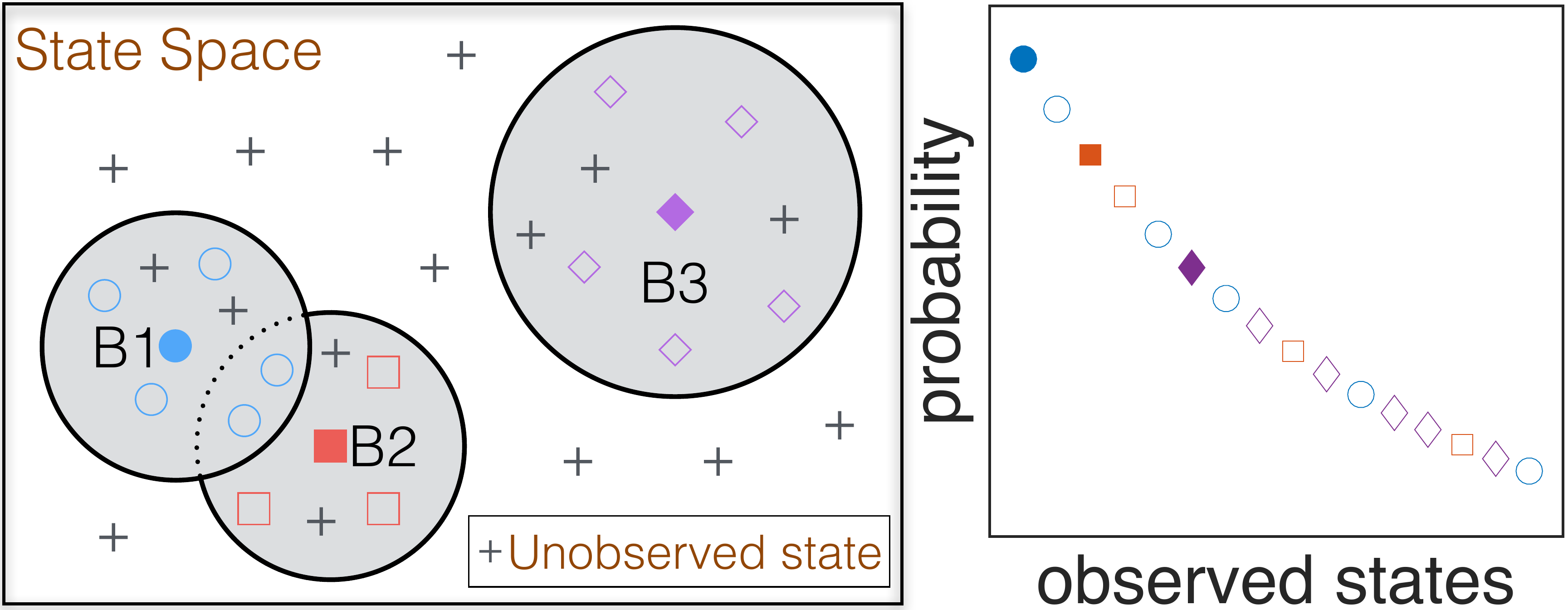} 
\caption{Schematic for the coarse-graining procedure. The filled symbols are bubble centers. The bubble structure is formed as follows: (1) Pick one set of experimental samples and sort the observed states in descending probability. (2) Consider only states that are not included in previous bubbles. Select a bubble center as the state with the highest probability (if multiple states attain the same highest probability, choose any of them as the bubble center). Form a new bubble by enclosing states with $L_{1}$ distance smaller than a cutoff radius to this bubble center (the cutoff radius may be increased for subsequent bubbles to ensure each one has comparable sample size). (3) The bubble structure is established with corresponding centers after all observed states are included. (4) Other samples and theoretical distributions are coarse grained with the same bubble structure.}
\label{Fig:schematics}
\end{figure}

With a proper coarse-graining procedure, we show, however, that a reliable comparison between two given output samples is achievable. There are a few factors that a reasonable coarse-graining procedure should consider. First, it should be constructed from experimental samples. On a large scale, classical simulation is no longer feasible. Experiments may nevertheless pick out the important output states with higher probabilities. Second, the procedure should be scalable: not only the measured events but all possible events should be grouped into some bubbles, where the number of bubbles should not be subject to the exponential growth. Third, the filtered information should still carry some knowledge of the full correlations in the outputs. This is because the essence of the many-body interference lies in the many-mode correlations: previous works \cite{Mayer2011Counting, Tichy2014Stringent, Tichy2014Interference} have shown that few-particle observables may not capture the full bosonic features and may falsely accept some fraudulent devices. Our proposed scheme takes the above factors into consideration, and coarse grain the states based on the $L_{1}$ distance measure. The $L_{1}$ distance between two occupation-number-basis states is defined as $L_{1} = \sum_{i}^{M} |\psi_{i} - \phi_{i}|$, where $\psi_{i}$ ($\phi_{i}$) is the occupation number in the $i$th mode for the state $\ket{\psi}$ ($\ket{\phi}$). Details of the coarse-graining procedure are shown in Fig.~\ref{Fig:schematics}. We note that this scheme is not the only way to perform coarse graining. We expect that other procedures meeting the above considerations may also work. Below, we show that useful and robust structures can be extracted from the samples using our method. 

\begin{figure*}[t!]
\hspace{-0.2cm}
\includegraphics[trim=1.8cm 0.2cm 4cm 1cm, clip,width=0.5\textwidth]{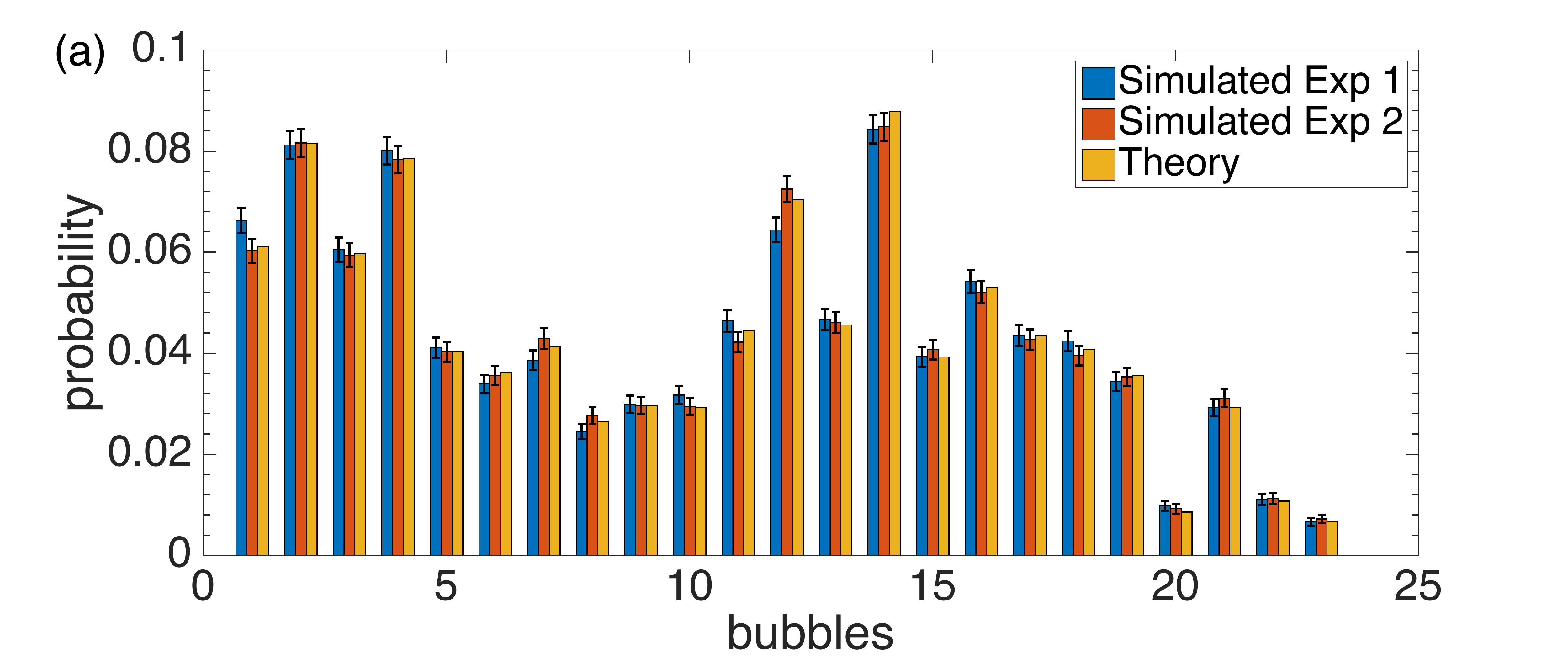} 
\hspace{-0.2cm}
\includegraphics[trim=1.8cm 0.2cm 4cm 1cm, clip,width=0.5\textwidth]{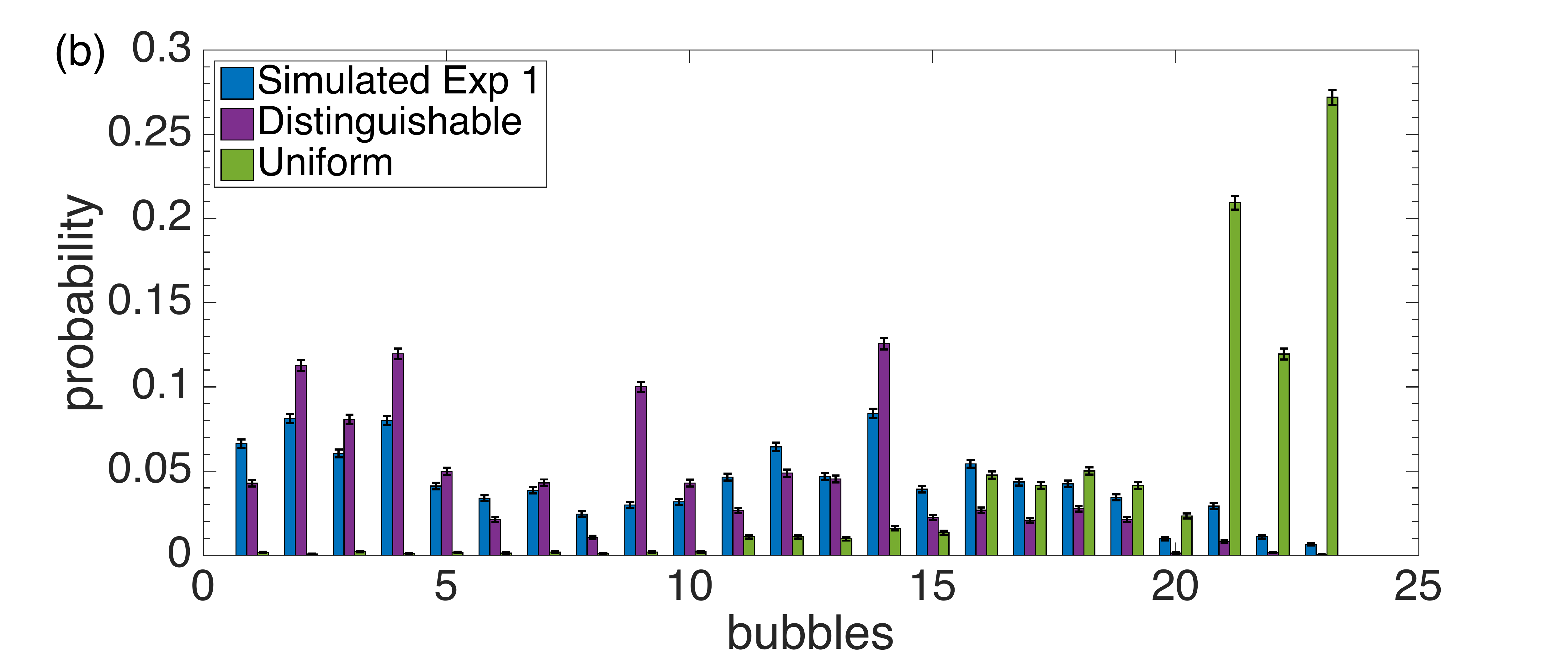} \\
\hspace{-0.2cm}
\includegraphics[trim=1.8cm 0.2cm 4cm 1cm, clip,width=0.5\textwidth]{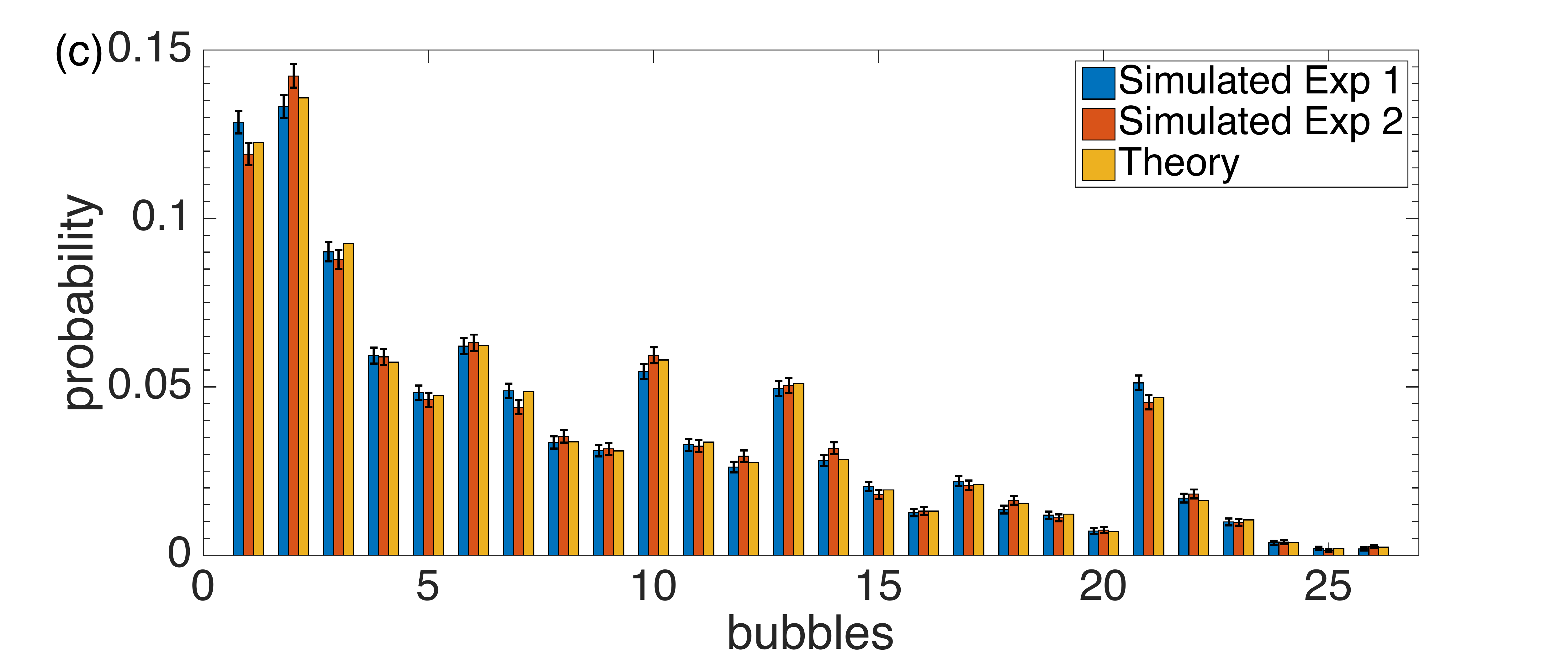}
\hspace{-0.2cm}
\includegraphics[trim=1.8cm 0.2cm 4cm 1cm, clip,width=0.5\textwidth]{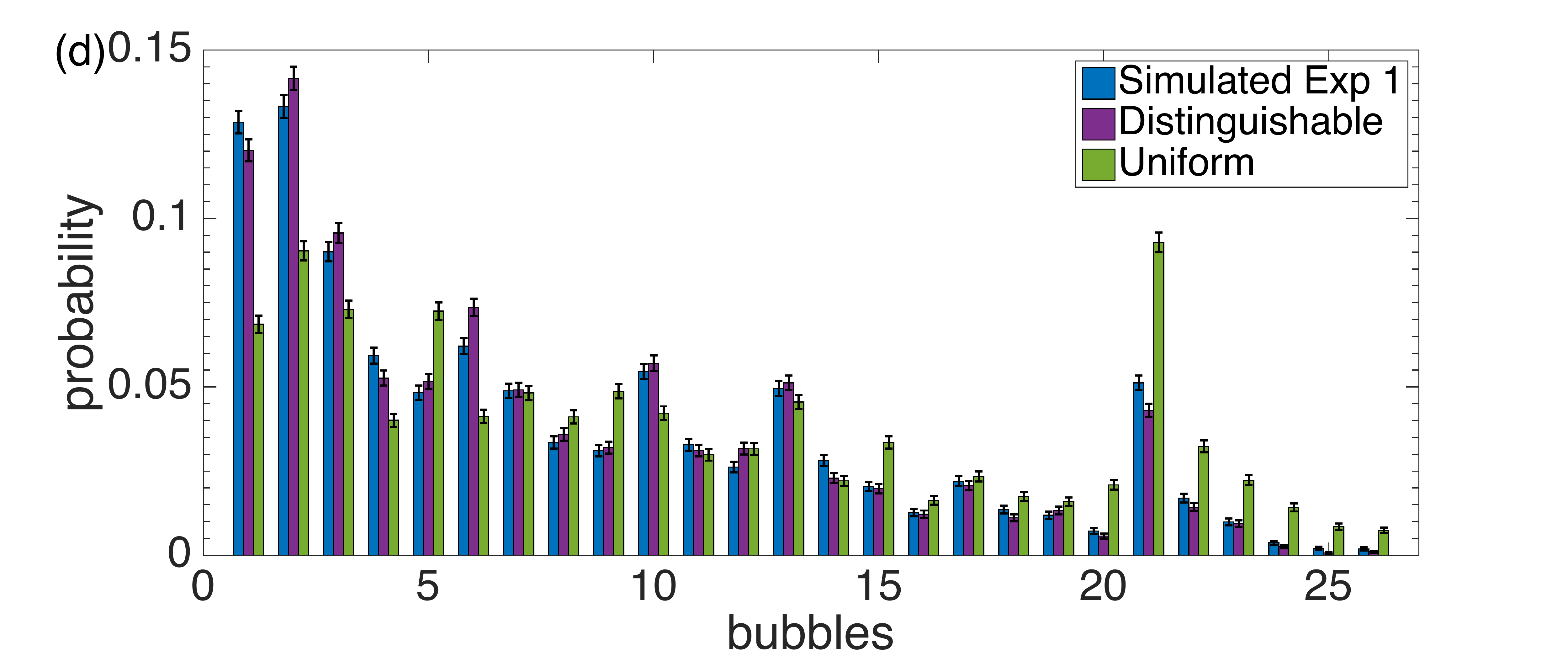}
\caption{Coarse-grained probability distributions. All samples have a sample size $N_{m}=10000$. The simulated experimental samples $\big( P^{Q}_{S1} \text{ and } P^{Q}_{S2} \big)$ are drawn from the boson sampler ($P^{Q}$); distinguishable ($P^{C}_{S}$) and uniform $\left(P^{U}_{S} \right)$  samples are drawn respectively from the classical ($P^{C}$) and uniform ($P^{U}$) samplers. Errors for the probabilities follow the standard deviations of the multinomial distribution. (a) and (b): probability distributions for the trapped-ion system with intermediate-time dynamics. The simulated system has $N=12$ phonons in a $M=12$ ion chain with one phonon on each ion as the input state. (c) and (d): probability distributions after a Haar-distributed random unitary transformation. The simulated system has $N=5$ particles in $M=40$ modes with $\ket{1,1,1,1,1, 0, \cdots \!, 0}$ as the input state.}
\label{Fig:distributions}
\end{figure*}

We simulate the experimental certification process with two different systems, one with trapped ions and one with Haar-distributed random unitaries. For all simulations, we choose a fixed sample size of $N_{m} = 10000$, which is a reasonable detection count in experiments, but is nevertheless smaller than $1\%$ of the Hilbert space dimension in study. 

For trapped ions, the transverse local phonons are used as indistinguishable bosons with the Hamiltonian given by \cite{Porras2004Bose, Zhu2006Trapped, Shen2014Scalable}   
\begin{equation}
H_{c} = \sum_{i}^{M} \hbar w_{x,i} a_{i}^{\dagger}a_{i} + \sum_{i< j}^{M} \hbar t_{ij} \left(a_{i}^{\dagger} a_{j} + a_{j}^{\dagger} a_{i} \right),
\label{Eq:IonHam}
\end{equation}
where $w_{x,i}=-\sum_{j \neq i}^{M} t_{0}/ |z_{i0}-z_{j0}|^{3}$, $t_{ij} = t_{0} / |z_{i0}-z_{j0}|^{3}$, and $t_{0} = e^{2}/(8\pi \epsilon_{0}m\omega_{x})$. $z_{i0}$ denotes the axial equilibrium position of the $i$th ion with mass $m$ and charge $e$. $w_{z}$ ($w_{x}$) is the axial (transverse) trapping frequency. In the simulation, we use experimentally relevant parameters $\omega_{z} = 2\pi \times 0.03 \,$MHz and $\omega_{x} = 2\pi \times 4 \,$MHz for $^{171}\text{Yb}^{+}$ ions and consider $N=12$ phonons on $M=12$ ions. The total Hilbert space size is $D=1352078$. The evolution time is chosen at some intermediate time ($\tau = 100 \, \mu$s) with interesting many-body dynamics (see Supplemental Material \cite{BosonSampling:Supp} for further results in the long time limit). For the random unitary process, we simulate $N=5$ particles in $M=40$ modes, a setting comparable to the current experimental regime with linear optics \cite{Broome2013Photonic, Spring2013Boson,Tillmann2013Experimental, Crespi2013Integrated, Spagnolo2013General, Spagnolo2014Experimental, Carolan2014On, Bentivegna2015Experimental} and in a limit where $M> N^{2}$ to suppress collision events. A Haar-distributed $ M \times M$ random unitary matrix is used, with a Hilbert space dimension of $D=1086008$.

Based on the coarse-graining procedure outlined in Fig.~\ref{Fig:schematics}, we group the sample events into different bubbles according to one set of experimental samples. The theoretical distributions are also subject to the same bubble structure. Fig.~\ref{Fig:distributions} shows an example of the coarse-grained distributions. This extracted information is robust and reliable with small sampling errors. By visual comparison, we can see that the boson sampling data $P^{Q}_{S1}$ and $P^{Q}_{S2}$ match closely with each other, but differ significantly from samples drawn from alternative distributions, such as the distinguishable ($P^{C}_{S}$) and uniform ($P^{U}_{S}$) samples~\cite{BosonSampling:Supp}. Here, it is also possible to compare the simulated data with the theoretical distribution $P^{Q}$, which will not be directly obtainable when experiments surpass classical simulation capabilities. By repeating the procedure, we observe that the comparisons are not sensitive to details of the coarse-graining method, such as the particular sample used to initiate the bubble structure. 

To quantify the degrees of matching between coarse-grained distributions, we employ the two-sample $\chi^{2}$ test. Under the null hypothesis wherein two samples are drawn from the same distribution, the $\chi^{2}$ statistics follow the $\chi^{2}$-distribution with degrees of freedom (df) equal to the number of bubbles ($N_{B}$) minus one. If the $\chi^{2}$ statistic is large with a small associated p-value, one nominally rejects the null hypothesis. Therefore, during the certification process, one should ideally accept the null if the pair of samples come from the same boson sampling device and reject it if one is sampled from an alternative distribution. Two types of errors can be incurred, a type I error (false positive) related to falsely rejecting the true null hypothesis and a type II error (false negative) associated with the failure to reject a false null hypothesis. Prior to the test, one sets a significance level $\alpha$, which will be the type I error rate if both samples are from the same distribution. 

\begin{figure}[t]
\includegraphics[trim=0cm 0cm 0cm 0cm, clip,width=0.95\columnwidth]{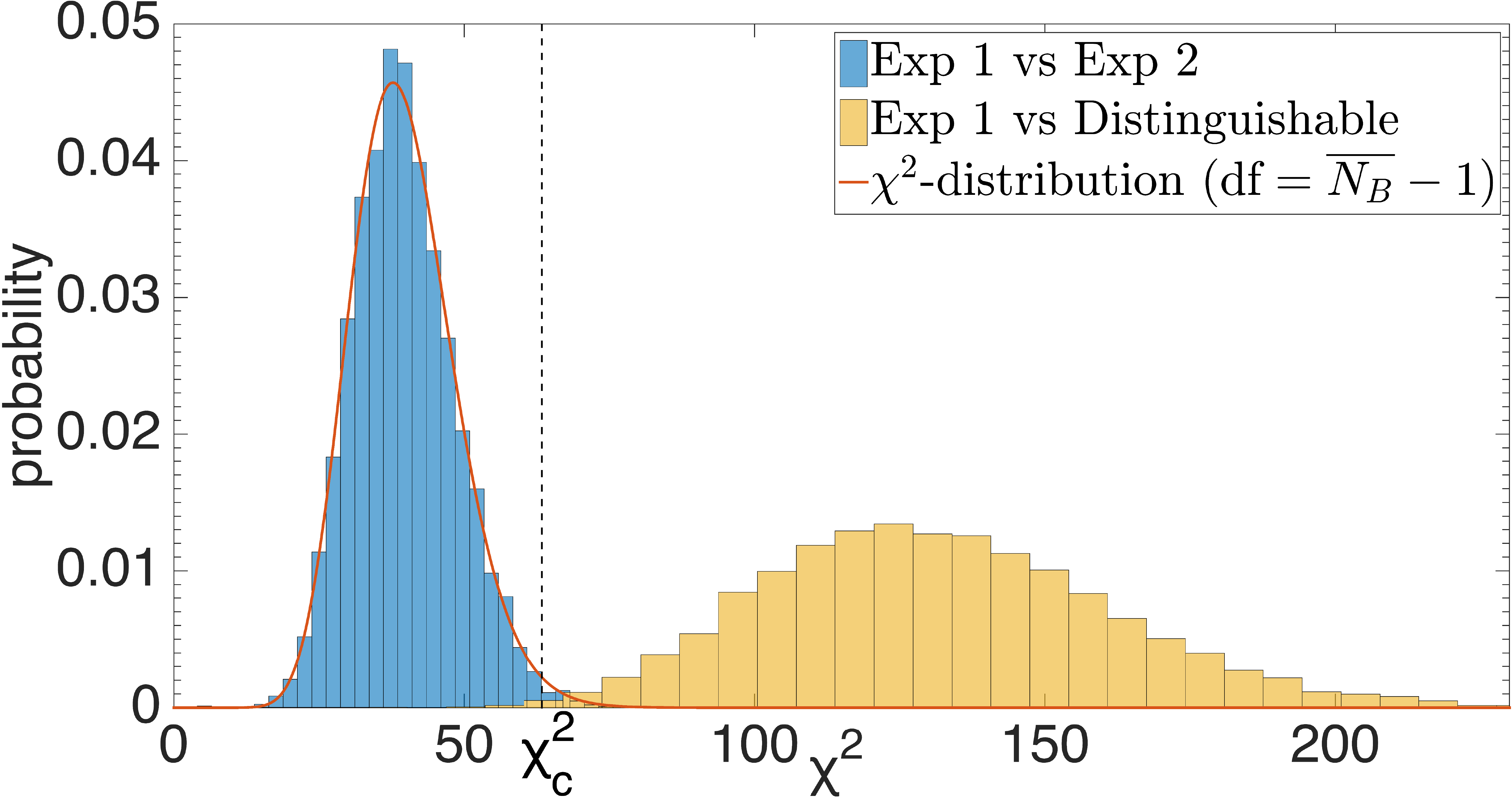} 
\caption{Distributions of the two-sample $\chi^{2}$ test-statistics for the random unitary process with $\overbar{N_{B}} \approx 40.8$. $N_{s}=10000$ sets of samples are generated in the simulation and distributions of the $\chi^{2}$ statistics between the corresponding pairs are plotted. The solid curve is the $\chi^{2}$-distribution with $\text{df}=\overbar{N_{B}}-1$. The dashed line marks the cutoff $\chi^{2}$ value at $\alpha = 1\%$ \cite{BosonSampling:Comment1}.}
\label{Fig:chi2dist}
\end{figure}

In simulation, we could generate many sets of samples and repeat the certification process to better gauge the error rates. Each time, we grouped the sample events into bubbles and compared one set of boson sampling data $P^{Q}_{S1}$ with various other samples ($P^{Q}_{S2}$, $P^{C}_{S}$, and $P^{U}_{S}$). A $\chi^{2}$ statistic and a p-value were computed for each $\chi^{2}$ test. This process was repeated for $N_{s} =10000$ runs, with the distributions of the $\chi^{2}$ statistics and p-values recorded. Fig.~\ref{Fig:chi2dist} presents the distributions of $\chi^{2}$ statistics for the random unitary process with the average number of bubbles $\overbar{N_{B}} \approx 40.8$. It can be seen clearly that the test statistics between $P^{Q}_{S1}$ and $P^{Q}_{S2}$ follow the $\chi^{2}$ distribution with $\text{df} = \overbar{N_{B}} -1$, whereas those between $P^{Q}_{S1}$ and $P^{C}_{S}$ fall on the far tail of the $\chi^{2}$ distribution, therefore offering a definitive answer regarding whether the samples are from the same distribution. More quantitatively, we calculate the pass rate $R$ at a given significance level $\alpha = 1\%$. Specifically, if the p-value is greater than $\alpha$, the comparison passes the test. Out of $N_{s}$ tests, the pass rates in percentage are reported in Table \ref{Table:passRate}. Between two sets of boson sampling data $P^{Q}_{S1}$ and $P^{Q}_{S2}$, the pass rates $R \approx 99\%$ for all cases, with the type I error rates $1-R$ being very close to $\alpha$ as expected. Between $P^{Q}_{S1}$ and alternative samples ($P^{C}_{S}$ and $P^{U}_{S}$), the pass rates reflect type II error. They can reach a few percent for smaller $N_{B}$ but drop to $<1\%$ as $N_{B}$ increases. This also presents a tradeoff between the information obscured by the sampling noise without coarse-graining and the information one discards by heavy coarse-graining. In general, to reduce sampling noise, each bubble should have a minimum number of observed events. For the $\chi^{2}$ test to work reliably, this minimum number is conventionally chosen to be 10. For a detection count of $N_{m} =10000$, a range of $ 20 \lesssim  N_{B}  \lesssim 100$ works well, with the requirement that the smallest bubble includes no less than 10 events (one could group together very small bubbles). Noticeably, this only depends on the number of detection counts, and does not scale up with the Hilbert space dimension. Our simulations demonstrate, in particular, that one could conclusively certify the boson sampling device with number of measurements less than $1\%$ of the Hilbert space dimension. 

\begin{table}[t]
\caption{The pass rates $R$ between simulated experimental sample 1 and various other samples. The two-sample $\chi^{2}$ test is performed to assess whether they come from the same distribution. The significance level $\alpha$ is set at $1\%$. For each pair of generated samples, if the p-value is greater than $\alpha$, the comparison passes the test. This is repeated for $N_{s} = 10000$ runs and pass rates are recorded. Noisy samples for the trapped-ion system include a $1\%$ ($3\%$) timing error, whereas a $1\%$ ($3\%$) random error is included in the random unitary matrix.}
\label{Table:passRate}%


\begin{ruledtabular}
\begin{tabular}{ccccccc}
 \multicolumn{7}{c}{pass rate ($\%$) compared to experimental sample 1} \\ 
\cline{1-7} 
& \multicolumn{3}{c}{Trapped Ions} & \multicolumn{3}{c}{Random Unitary} \\
\cline{1-1}  \cline{2-4}  \cline{5-7} 
$\overbar{N_{B}}$ & $24.2$  & $40.5$ & $69.9$ & $25.9$ & $40.8$ & $70.5$ \\ \hline
Exp 2 & $99.1$  & $99.2$ & $99.1$ & $ 99.1 $ & $ 99.2 $ & $ 99.1 $ \\
Exp 2 (1\% Noise)  & $98.0 $  & $98.2$ & 98.2 & $98.9$ & $98.9$ & $99.1$ \\
Exp 2 (3\% Noise)  & $71.1$  & $77.1$ & 75.9   & $98.3$ & $98.4$ & $98.6$ \\
Distinguishable & $0 $  & $0 $ & $0 $  & $ 6.7 $ & $0.3$ & $0.03$ \\
Uniform & $0 $  & $0$ & $0$ & $0$ & $0$ & $0$ \\
\end{tabular}
\end{ruledtabular}

\end{table}

Furthermore, we consider the effects of realistic noise in experiments. In the case of trapped ions, we included a $1\%$ $(3\%)$ systematic error in the timing ($1 \, \mu$s ($3 \, \mu$s) shift in $\tau$). This is also equivalent to a $1\%$ ($3\%$) error in the hopping amplitude $t_{ij} \sim \omega_{z}^{2} / \omega_{x}$ \cite{Shen2014Scalable}, which translates to a respective shift in the trapping frequencies in experiments. For the random unitary process, we added $1\%$ $(3\%)$ random noise to the unitary matrix (see Supplemental Material for details \cite{BosonSampling:Supp}). As seen in Table \ref{Table:passRate}, with small noise ($\sim 1\%$) the type I error rates are kept in check ($\lesssim 2\%$). When the noise becomes substantial, pass rates may drop sharply if the unitary process changes considerably. This also shows the sensitivity of our method to strong noise and it could serve as a stringent certification test. The dissimilar sensitivity to noise for the two systems is due to the different ways noise is included and natures of the noise (systematic versus random).

In conclusion, we have shown that useful and robust information can be extracted from the coarse-grained measurements for boson sampling. The coarse-grained distributions can be further used to certify the boson sampling device. We expect this method to be handy when experiments progress beyond classical capabilities. It should also be a useful tool for other generic many-body certification problems. 

\begin{acknowledgments}
This work is supported by the ARL, the IARPA LogiQ program, and the AFOSR MURI program.
\end{acknowledgments}


\begin{thebibliography}{30}
\bibitem [{Bos({\natexlab{b}})}]{BosonSampling:Comment1}%
  \BibitemOpen
  \href@noop {} {}\bibinfo {howpublished} {Rigorously speaking, each set of
  generated samples may have different $N_{B}$ since we do not strictly control
  it. Hence each $\chi^{2}$ statistic may have different df. However, $N_{B}$
  does not vary much in each run, and we can see that the $\chi^{2}$ statistics
  for a pair of experimental samples roughly follow the $\chi^{2}$-distribution
  with $\text{df}=\overbar{N_{B}}-1$. The pass rates in table
  \ref{Table:passRate} are calculated properly as p-values are computed for
  each pair of samples with the actual df.}
\end{thebibliography}

%

\onecolumngrid

\newpage

\section{Supplemental Material: Certification of Boson Sampling Devices with Coarse-Grained Measurements}

\begin{quote}
In this Supplemental Material, we include additional results for the trapped-ion system with long-time dynamics and for boson sampling data with experimental noise. Further analysis for the two-sample $\chi^{2}$ test is provided. 
\end{quote}

\section{Results for the Long-time trapped-ion system}

\begin{figure}[h]
\hspace{-0.2cm}
\includegraphics[trim=1.8cm 0.2cm 4cm 1cm, clip,width=0.5\textwidth]{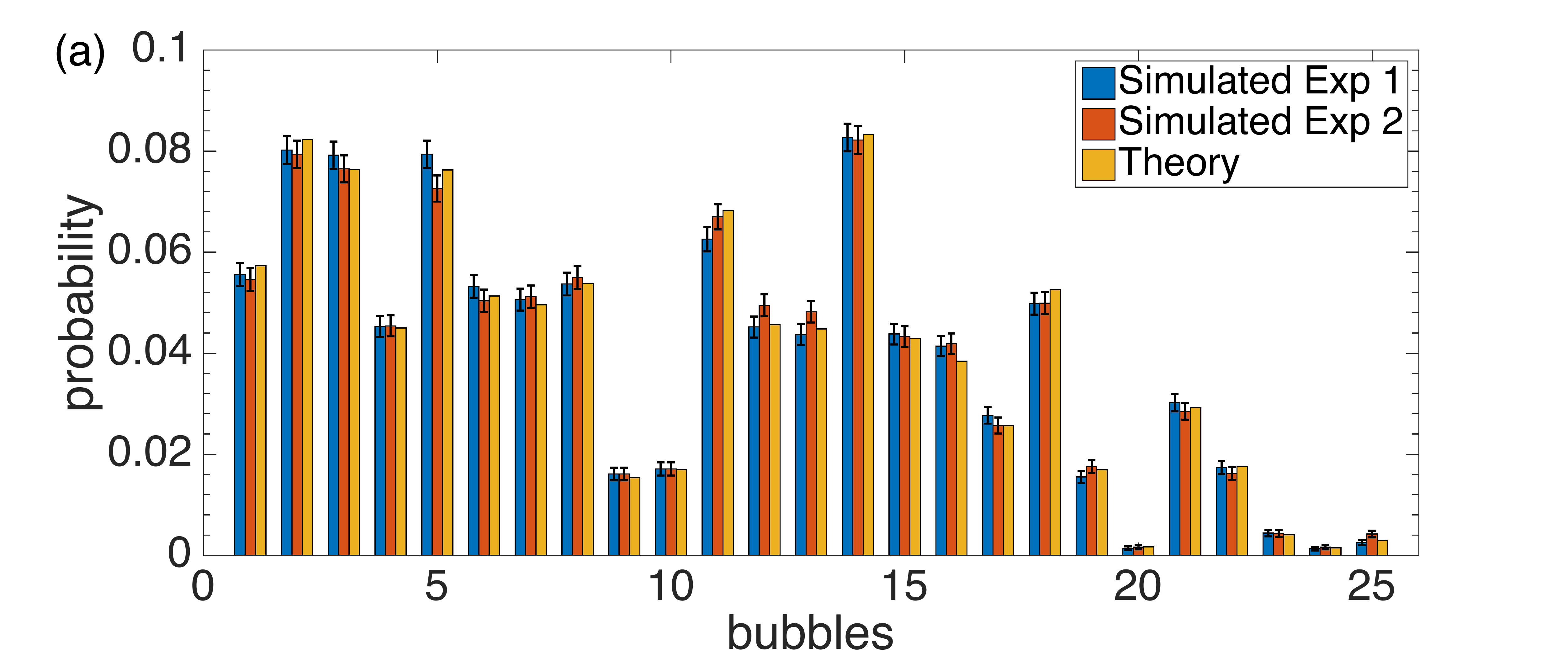} 
\hspace{-0.2cm}
\includegraphics[trim=1.8cm 0.2cm 4cm 1cm, clip,width=0.5\textwidth]{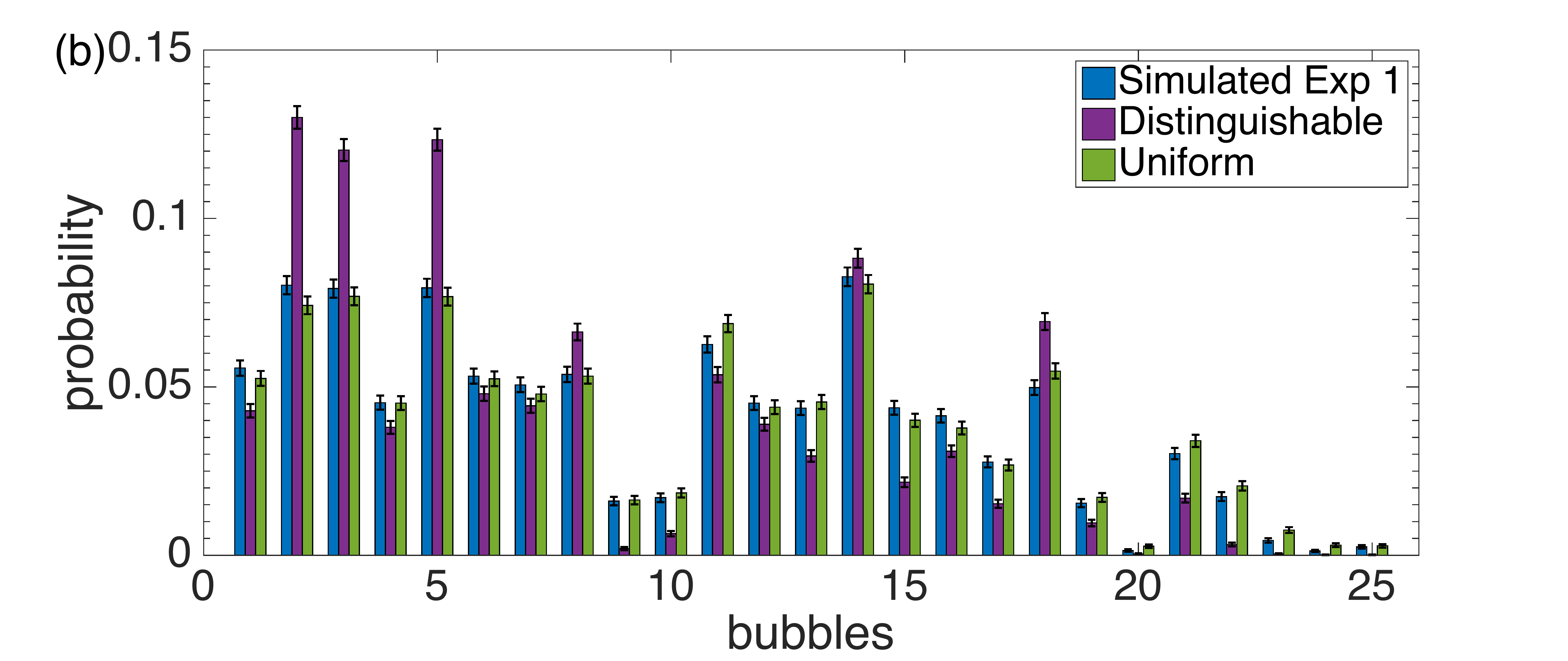} \\
\caption{Coarse-grained probability distributions for the trapped-ion system with long-time dynamics. All samples have a sample size $N_{m}=10000$. The simulated experimental samples $\big( P^{Q}_{S1} \text{ and } P^{Q}_{S2} \big)$ are drawn from the boson sampler ($P^{Q}$); distinguishable ($P^{C}_{S}$) and uniform $\left(P^{U}_{S} \right)$  samples are drawn respectively from the classical ($P^{C}$) and uniform ($P^{U}$) samplers. Errors for the probabilities follow the standard deviations of the multinomial distribution. The simulated system has $N=12$ phonons in a $M=12$ ion chain with one phonon on each ion as the input state.}
\label{Fig:ion2dist}
\end{figure}
 
In the main text, we included results for both the trapped-ion system with intermediate-time dynamics ($\tau = 100 \, \mu$s) and for the random unitary process (Fig.~3 and Table~I of the main text). In Fig.~\ref{Fig:ion2dist} and Table \ref{Table:passRateSup} here, we add the coarse-grained distributions and the pass rate  results for the long-time trapped-ion system (at $\tau = 10\,$ms). With these coarse-grained distributions, we would like to discuss some differences between the trapped-ion system and the random unitary process. In the prototypical boson sampling problem, the hardness-of-simulation argument is based on randomly selected unitaries \cite{Aaronson2011Computational}. This is because for some structured unitary process, fast approximation algorithms may exist. For the trapped-ion system, the phonon normal modes are fixed (fixed time-independent Hamiltonian in Eq.~1 of the main text), so we may be able to extract some distinctive structures from the output probability distributions. Nevertheless, we still expect the complex many-body dynamics in trapped ions to be classically intractable. 

\begin{table}[b]
\caption{The pass rates $R$ between simulated experimental sample 1 and various other samples for the trapped-ion system with long-time dynamics. The two-sample $\chi^{2}$ test is performed to assess whether they come from the same distribution. The significance level $\alpha$ is set at $1\%$. For each pair of generated samples, if the p-value is greater than $\alpha$, the comparison passes the test. This is repeated for $N_{s} = 10000$ runs and pass rates are recorded.}
\label{Table:passRateSup}%

\begin{tabular*}{0.45\textwidth}{@{\extracolsep{\fill}} cccc}
\hline \hline
\multicolumn{4}{c}{pass rate ($\%$) compared to experimental sample 1} \\ 
\hline 
& \multicolumn{3}{c}{Trapped Ions (Long time)} \\
\cline{1-1}  \cline{2-4} 
Average No.\ of bubbles & \hspace{.3cm} $24.3$  & \hspace{.3cm} $39.0$ &  $66.5$ \\ \hline
Experimental Sample 2 & \hspace{.3cm} $98.9$  & \hspace{.3cm} $99.0$ & $99.2$ \\
Distinguishable Sample & \hspace{.3cm} $0 $  & \hspace{.3cm} $0 $ & $0 $ \\
Uniform Sample & \hspace{.3cm} $5.7$  & \hspace{.3cm} $1.3$ & $0.2$ \\
\hline \hline
\end{tabular*}
\end{table}

From the coarse-grained distributions, we can see that the boson sampling data $P^{Q}_{S}$ for intermediate-time trapped ions are conspicuously different from alternative samples, such as the distinguishable sample $P^{C}_{S}$ and the uniform sample $P^{U}_{S}$. On the other hand, the boson sampling data bear resemblance to the distinguishable samples $P^{C}_{S}$ for random unitary processes, and the long-time trapped-ion phonon distributions are instead in closer proximity to the uniform samples $P^{U}_{S}$. The qualitative difference arises from the fact that the random unitary process has a Haar-distributed random unitary matrix whereas the long-time trapped-ion unitary only has random phases (eigenvalues) with a fixed crystal mode structure. We have further tested this idea by simulating the long-time trapped-ion dynamics with a unitary matrix given by the fixed mode structure and random phases. Certification results do support the notion that this system produces output probabilities in closer proximity to the uniform samples (for example, the pass rate $R \approx 3\%$ between the simulated samples and uniform samples and $R \approx 0\%$ between the simulated samples and distinguishable samples for $\overbar{N_{B}} \approx 39.1$). This in addition shows our method is sensitive to some structures hidden in the many-body interference process.

\section{Noisy Samples}

\begin{figure}[t]
\hspace{-0.2cm}
\includegraphics[trim=1.8cm 0.2cm 4cm 1cm, clip,width=0.5\textwidth]{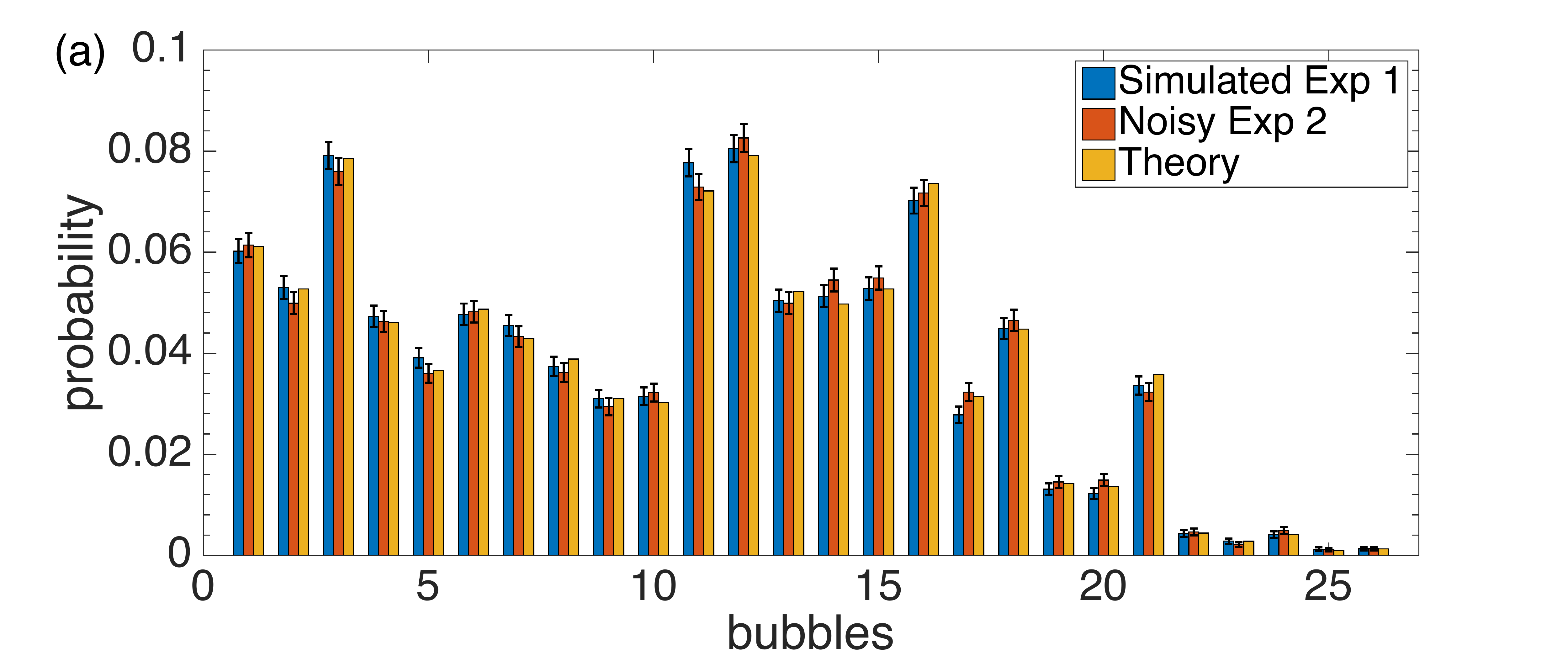} 
\hspace{-0.2cm}
\includegraphics[trim=1.8cm 0.2cm 4cm 1cm, clip,width=0.5\textwidth]{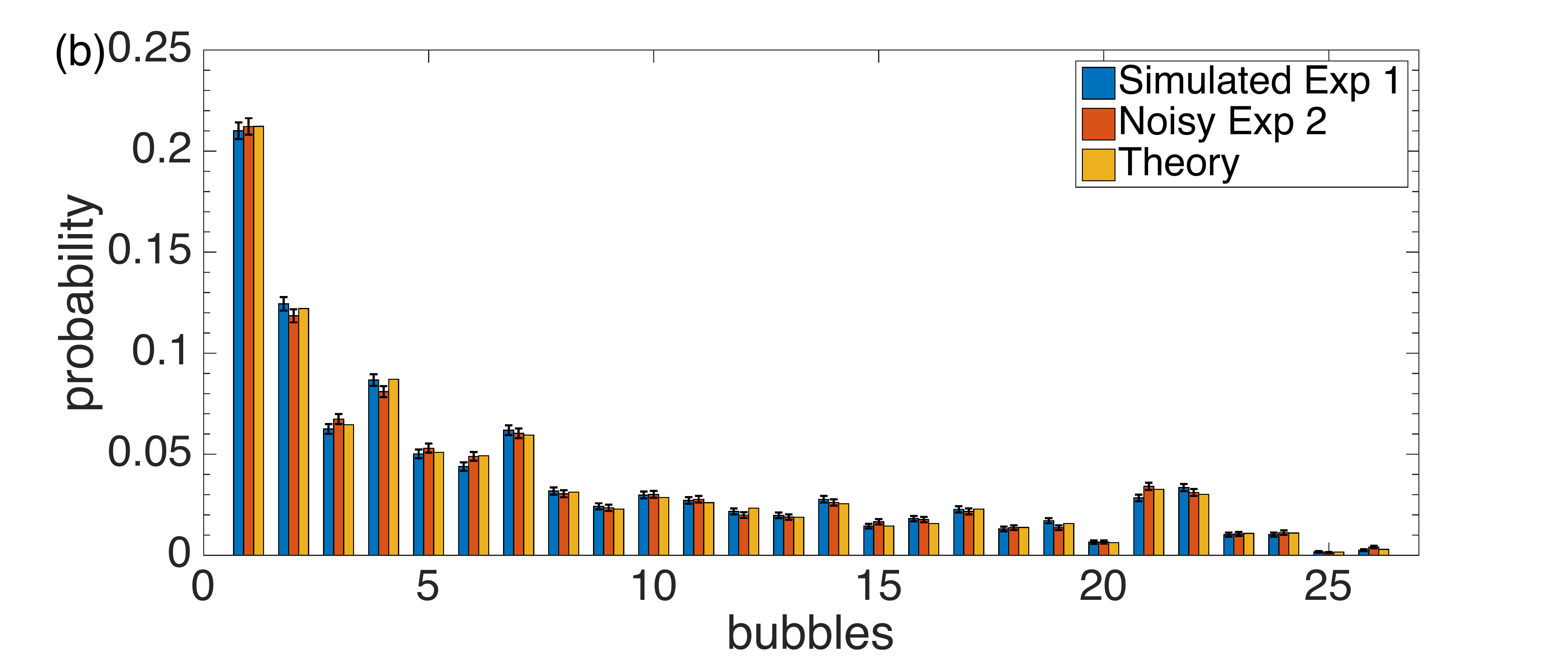} \\
\caption{Coarse-grained probability distributions for the noisy samples. (a) 
Trapped-ion system of intermediate-time dynamics with a $1\%$ timing error included in the noisy sample. The simulated system has $N=12$ phonons in a $M=12$ ion chain with one phonon on each ion as the input state. (b) Distributions after a random unitary transformation. A $1\%$ random noise is added to the unitary process in the noisy sample. The simulated system has $N=5$ particles in $M=40$ modes with $\ket{1,1,1,1,1, 0, \cdots \!, 0}$ as the input state.}
\label{Fig:noisyDist}
\end{figure}

Fig.~\ref{Fig:noisyDist} presents some coarse-grained distributions including the noisy samples. Visually, the $1\%$ noise in the unitaries does not lead to substantial changes to the distributions. The noise we include for the intermediate-time trapped-ion system is a $1\%$ systematic error in the total time (shift from $\tau = 100 \, \mu$s to $\tau = 101 \, \mu$s). For the random unitary process, in order to preserve the unitarity of the process, we first find the effective Hamiltonian for the Haar-distributed random unitary process, and subsequently add a $1\%$ random noise to each entry of the hermitian Hamiltonian matrix.

\begin{figure}[b]
\includegraphics[trim=0cm 0cm 0cm 0cm, clip,width=0.45\columnwidth]{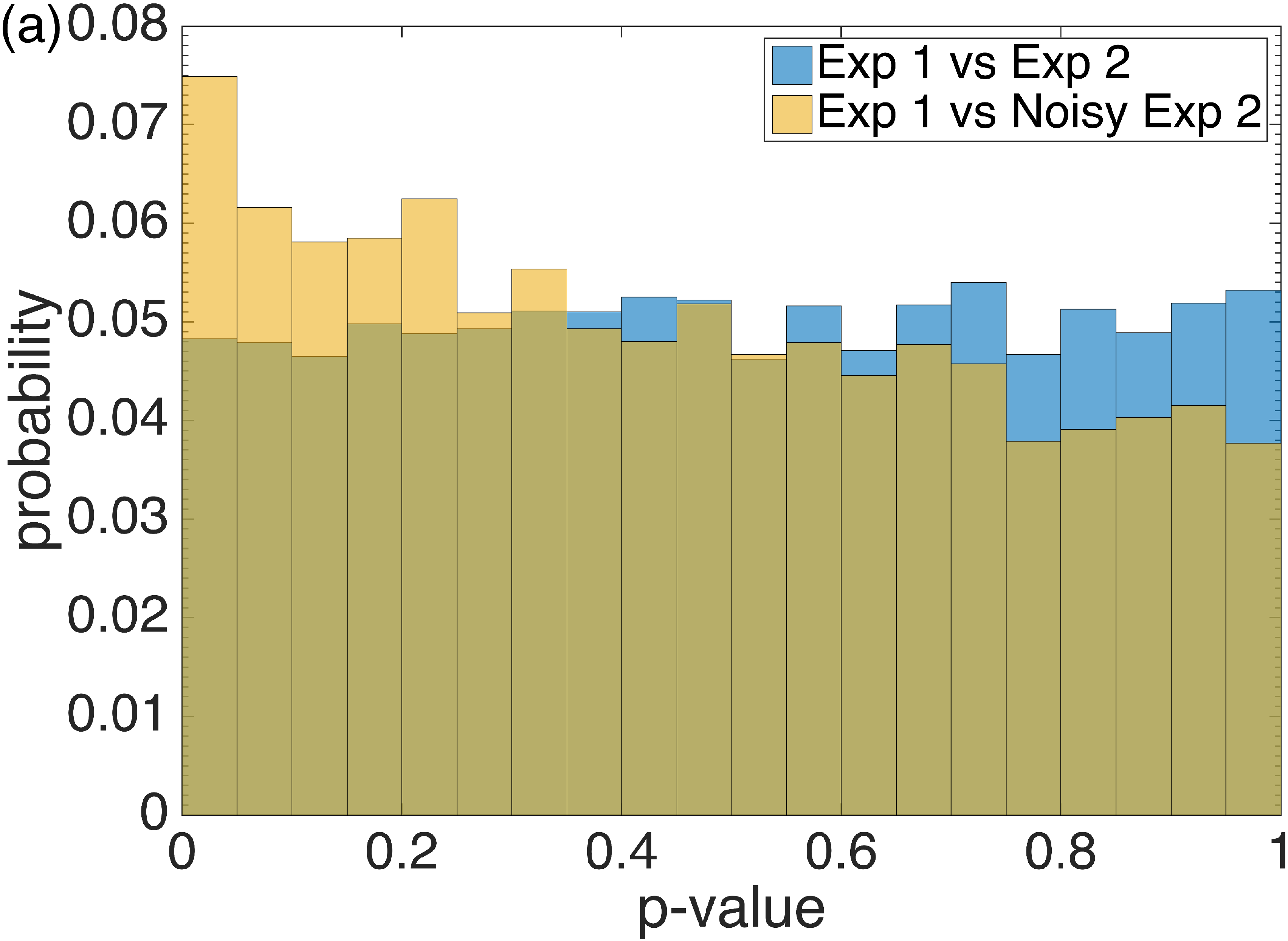} 
\includegraphics[trim=0cm 0cm 0cm 0cm, clip,width=0.45\columnwidth]{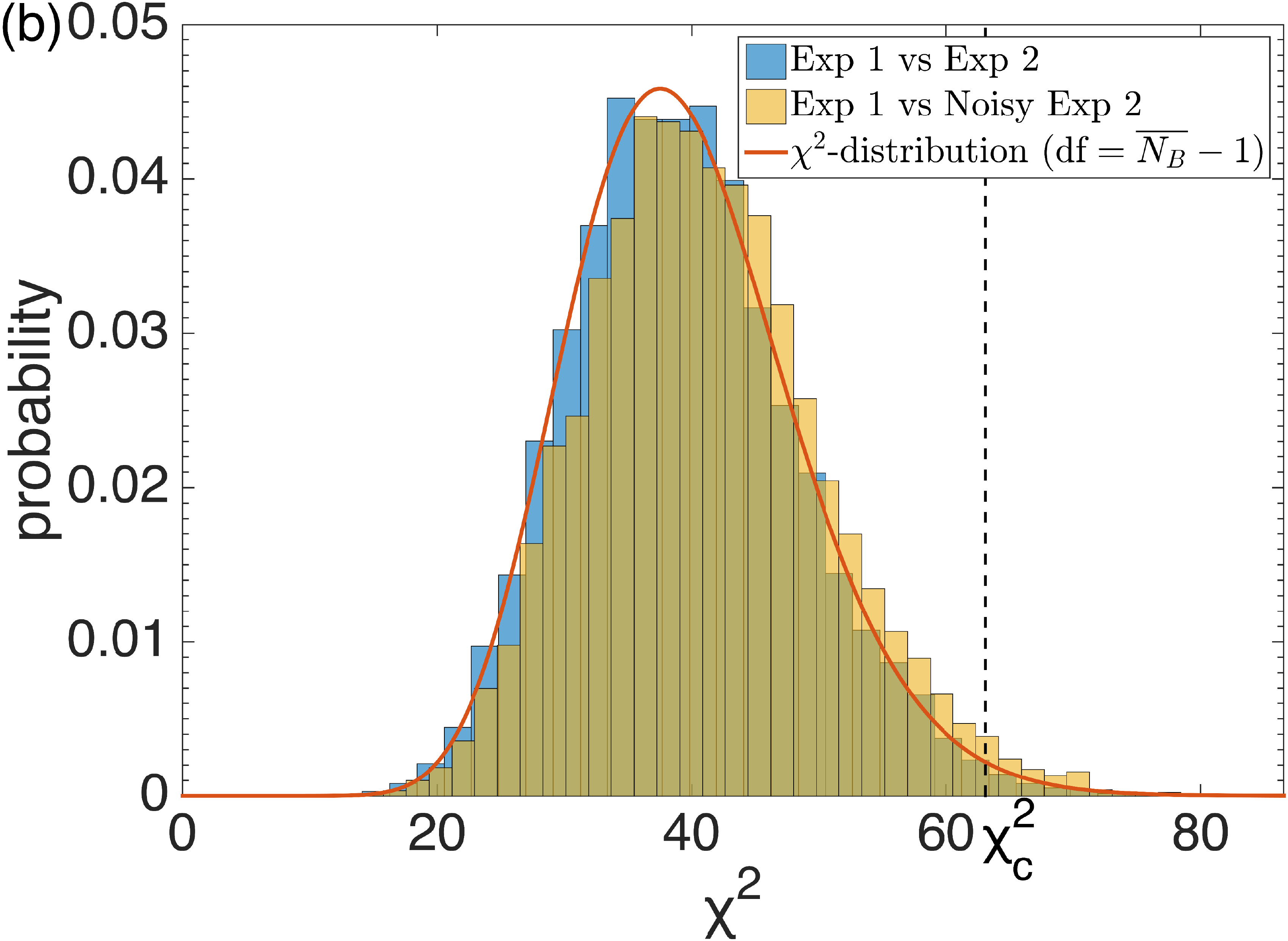} 
\caption{Distributions of p-values and the two-sample $\chi^{2}$ test-statistics for the trapped-ion system with intermediate-time dynamics and $\overbar{N_{B}} \approx 40.5$. A $\chi^{2}$ statistic and a p-value are calculated for each pair of samples, with a total $N_{s}=10000$ runs generated in the simulation. (a) If two samples come from the same distribution, p-values should be uniformly distributed in $[0,1]$. (b) The solid curve is the $\chi^{2}$-distribution with $\overbar{N_{B}}-1$ degrees of freedom. The dashed line marks the cutoff $\chi^{2}$ value at $1\%$ significance level. }
\label{Fig:pvalues}
\end{figure}

\section{Further analysis on two-sample $\chi^{2}$ test}

For the two-sample $\chi^{2}$ test, if the pair of samples comes from the same distribution, the $\chi^{2}$ statistics should follow the $\chi^{2}$-distribution with degrees of freedom being the number of bubbles minus one. Also, if the null hypothesis is correct, p-values should be uniformly distributed in $[0,1]$. In the main text, we plotted the distributions of the $\chi^{2}$ statistics between two boson sampling data and between one boson sampling data and a distinguishable sample. One could clearly see the distinction in comparison. Here, we include in Fig.~\ref{Fig:pvalues} the plots for the noisy samples too, in the case of intermediate-time trapped-ion system with $\overbar{N_{B}} \approx 40.5$. Without noise in experiment, it is evident that the $\chi^{2}$ statistics follow the $\chi^{2}$-distribution and the p-values are almost uniformly distributed. With experimental noise, $\chi^{2}$ statistics shift to larger values, with p-values tilted towards smaller values. Therefore, type I error is going to increase with noise in experiments. Nevertheless, with small noise ($1\%$ in this case), one can still definitively and correctly conclude that these samples are equivalent with $\gtrsim 98\%$ confidence level (as tabled in the main text).

Other than the pass rates $R$ tabled in the main text, here we add a table for the p-values too, which offers additional information regarding the error rates without comparing to a specific significance level $\alpha$. The mean and standard deviation of p-values are reported in Table \ref{Table:pvalues} for a set of $N_{s}=10000$ runs. In the case of two boson sampling data, the mean and standard deviation are very close to the theoretical values $0.5$ and $0.289$ respectively. On the other hand, p-values are distinctively smaller against alternative samples. Some far-fetched distributions even have p-values smaller than the machine level ($<10^{-323}$) with the boson sampling data, which illustrates the effectiveness of our certification method. With an appropriate significance level $\alpha$ and suitable number of bubbles $N_{B}$, one could minimize both type I and type II errors in the experimental certification process.

\begin{table*}[t]
\caption{Mean and standard deviation of p-values for the two-sample $\chi^{2}$ test between one simulated experimental sample and various other samples. A p-value is calculated for each pair of generated samples and this is repeated for a total of $N_{s} = 10000$ sets. Mean and standard deviation (in brackets) of those p-values are tabled below. If two samples come from the same distribution, p-values should be uniformly distributed in $[0,1]$ with mean $0.5$ and standard deviation $0.289$. If not, p-values should be small. 
Noisy samples for the trapped-ion system include a $1\%$ timing error, whereas a $1\%$ random error is included in the random unitary matrix. A value of $0$ in the table indicates that the p-value is extremely small ($<10^{-323}$ at the machine level).}
\label{Table:pvalues}%
\begin{ruledtabular}
\begin{tabular}{cccccccccc}
 & \multicolumn{9}{c}{Simulated experimental sample 1 (p-values)} \\ 
\cline{2-10} 
& \multicolumn{3}{c}{Trapped Ions (Int.\ time)} & \multicolumn{3}{c}{Trapped Ions (Long time)} & \multicolumn{3}{c}{Random Unitary} \\
\cline{1-1}  \cline{2-4}  \cline{5-7} \cline{8-10}    
Average No.\ of bubbles & $24.2$  & $40.5$ & $69.9$  & $24.3$ & $39.0$  &  $66.5$ & $25.9$ & $40.8$ & $70.5$ \\ \hline
Experimental sample 2 & $0.50$  & $0.51$ & $0.51$  & $0.50$ & $ 0.50 $  &  $ 0.51 $ & $ 0.50 $ & $ 0.50 $ & $ 0.50 $ \\
			 & $ (0.29) $  & $(0.29)$ & $ (0.29) $  & $  (0.29) $ & $  (0.29) $  &  $ (0.29) $ & $ (0.29) $ & $ (0.29) $ & $ (0.29)$  \\
Noisy experimental sample 2 & $0.44 $  & $0.45$ & 0.44   & - & -  &  - & $0.50$ & $0.49$ & $0.50$ \\
			 & $(0.29) $  & $(0.29)$ & $(0.29)$  & - & -  &  - & $(0.29)$ & $(0.29)$ & $(0.29)$ \\
Distinguishable sample & $0 $  & $0 $ & $0 $  & $0 $ & $0 $  &  $0 $ & $ 0.0053 $ & $2.1 \! \times \! 10^{-4}$ & $7.9 \! \times \! 10^{-5}$ \\
			 & -   & - &  - &  -&   -& -  & $(0.033)$ & (0.0076) & $(0.0071)$ \\
Uniform sample & $0 $  & $0$ & $0$  & $0.0043$ & $8.4 \! \times \! 10^{-4}$  &  $1.1 \! \times \! 10^{-4}$ & $0$ & $0$ & $0$ \\
			 &-   &-  &  - & $(0.029)$ & $(0.0098)$  &  $(0.0027)$ & - & - & -  \\
\end{tabular}
\end{ruledtabular}
\end{table*}

\end{document}